# PREVENTION OF WORMHOLE ATTACK IN WIRELESS SENSOR NETWORK

Dhara Buch[1] and Devesh Jinwala[2]

[1]Department of Computer Engineering, Government Engineering College,Rajkot, Gujarat
dh.buch@gmail.com
[2]Department of Computer Engineering, S. V. National Institute of Technology, Surat, Gujarat
dcj@svnit.ac.in

***ABSTRACT***

*Ubiquitous and pervasive applications, where the Wireless Sensor Networks are typically deployed, lead to the susceptibility to many kinds of security attacks. Sensors used for real time response capability also make it difficult to devise the resource intensive security protocols because of their limited battery, power, memory and processing capabilities. One of potent form of Denial of Service attacks is Wormhole attack that affects on the network layer. In this paper, the techniques dealing with wormhole attack are investigated and an approach for wormhole prevention is proposed. Our approach is based on the analysis of the two-hop neighbors forwarding Route Reply packet. To check the validity of the sender, a unique key between the individual sensor node and the base station is required to be generated by suitable scheme.*

## 1.INTRODUCTION

Deployment of Wireless Sensor Network is mainly in hostile environments like military battle field, habitat monitoring, nuclear power plants, target tracking, seismic monitoring, fire and flood detection etc. where constant monitoring and real time response are of pioneer requirement. Loss of normal messages can be allowed by such applications, but they cannot tolerate the loss of numerous packets of critical event messages [1]. This kind of need makes the sensor nodes an essential part of the network. Inherently, a wireless sensor network is an interconnection among hundreds, thousands or millions of sensor nodes. A sensor node is an embedded device that integrates a number of microprocessor components onto a single chip. Although a sensor node is capable of sensing, data processing and communication tasks, their limited memory capacity, limited battery power, low bandwidth and low computational power makesthe sensor network vulnerable to many kinds of attacks. [2]. Even unlike the wired networks where attackers are prevented by the physical media, the open nature of wireless medium makes it easy for outsider attackers to interfere and interrupt the legitimate traffic. [3] This leads to various security issues like key establishment, secrecy, authentication, privacy, secure routing etc.

Wormhole attack is one of the Denial-of-Service attacks effective on the network layer, that can affect network routing, data aggregation and location based wireless security. [3] The wormhole attack may be launched by a single or a pair of collaborating nodes. In commonly found two ended wormhole, one end overhears the packets and forwards them through the tunnel to the other end, where the packets are replayed to local area. In case when they only forward all the packets without altering the content, they are helping the network to accomplish transmission faster. But in majority of the cases, it either drops or selectively forwards the packets, leading to the network disruption. Wormhole attack does not require MAC protocol information as well as it is immune to cryptographic techniques. [4] This makes it very difficult to detect.

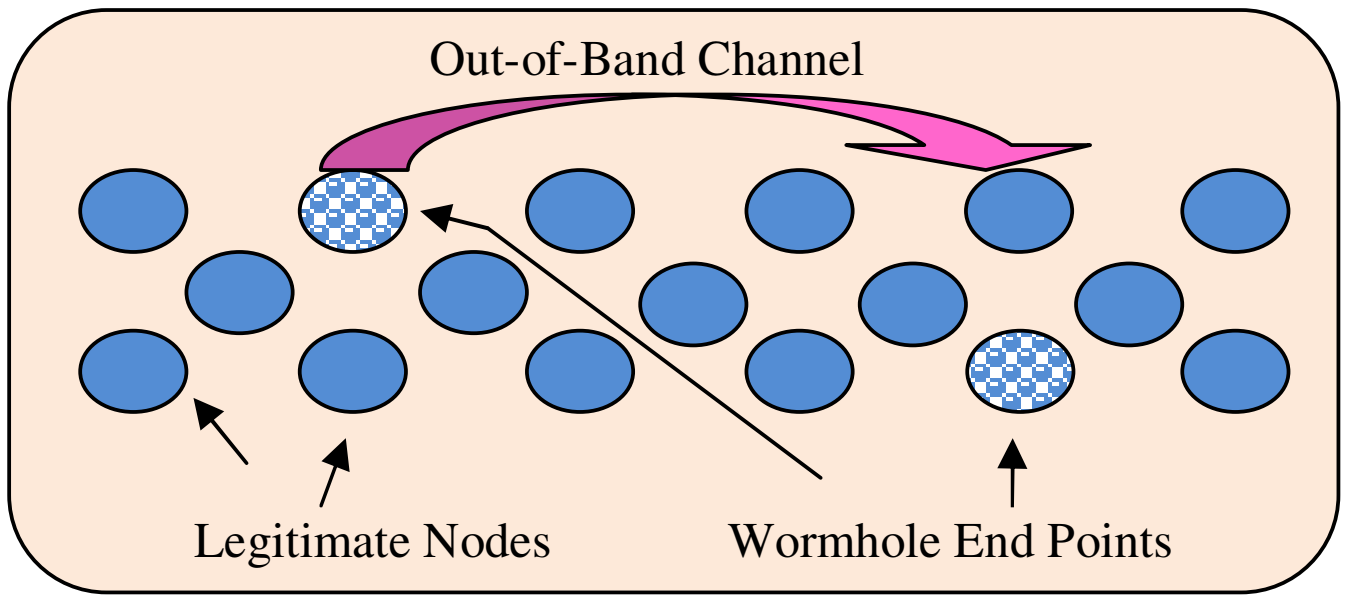

Figure 1. Two-Ended Wormhole Link

A number of approaches have been proposed for handling wormhole attack. Some approaches only detect the presence of wormhole in the network, while some approaches also focus on avoiding or preventing the wormhole attack. Majority of the techniques presented require additional hardware support, tight time synchronization, localization information or may be confined to specific routing algorithm.

An approach for preventing wormhole attack has been presented in this paper. No special hardware or time synchronization is required for this method. Additionally, only self geographical location is required for the proposed key generation phase. The mechanism implements conditional forwarding of Route Reply packet based on the validity of the two-hop neighbor forwarding it. The route is selected for transmission only if each node in reverse path till the source node validates the two-hop sender node.

The remaining sections of the paper are organized as follows: Section 2 discusses Denial of Service attack and gives an overview of various types and models of wormhole. Section 3 presents a brief review of the existing techniques for wormhole attack prevention. The proposed approach is explained in section 4. Simulation setup is given in section 5 while simulation results and their analysis are discussed section 6. Finally, conclusion is drawn and possible future work is proposed in section 7.

## 2. BACKGROUND

Effective performance in hostile area increases the application of wireless sensor network day by day, but the open nature of wireless medium and especially hardware limitations of a wireless sensor network themselves make it vulnerable to many kinds of security attacks. Adversary, with a large amount of power supply, memory and processing abilities and capacity for high-power radio transmission, leads to various kinds of attacks to the network [10]. This makes security a major issue to be concentrated well in a sensor network.

### 2.1.Threat Models

Participation of attackers, the way they affect network and their capacities classify adversaries into various models. Depending upon whether the attacker is the part of the network itself or not, categorizes them into *Outsider* and *Insider* attackers. *Passive* attackers do not alter the packet data but they only interfere in normal network traffic flow, while data is updated by *Active* attackers. *Mote* class attackers have limited power capacity to affect some sensor nodes only, while more powerful devices like laptop can also be victims of *Laptop* class attackers.

Many routing protocols are devised but almost none of them is designed with security as a goal[1]. Securing with shared symmetric keys also brings a question of distributing key, while it is too expensive to implement asymmetric cryptography. Even after handling all these issues if security is imposed, there is an attack namely, Denial of Service attack that can disturb, disrupt or can stop the legitimate routing operations even without the knowledge of the encryption. This feature makes it very important to identify and to defend against.

## 2.2. Denial of Service Attack

A Denial of Service (DoS) attack is an attempt to make a computer system (server or client) or some other resource unavailable to legitimate users. In general, it aims to prevent some services from functioning efficiently either temporarily or indefinitely.Hardware failures, environmental conditions, software bugs or resource exhaustion can lead to Denial of Service attacks. Various types of DoS attacks work at different layer and affect differently to the network, where one kind of DoS attack is wormhole attack.

## 2.3. Wormhole Attack Model

Wormhole attack is a network layer attack that can affect the network even without the knowledge of cryptographic techniques implemented. This is the reason why it is very difficult to detect. It is caused by one, two or more number of nodes. In most commonly type of two-ended wormhole, one end tunnels the packets via wormhole link and the other end, on receiving packets, replays them to the local area. The types and models of wormhole are explained here.

### 2.3.1. Types of Wormhole Attack

Number of nodes involved in establishing wormhole and the way to establish it classifies wormhole into the following types.

***Wormhole using Out-of-Band Channel***

In this two-ended wormhole, a dedicated out-of-band high bandwidth channel is placed between end points to create a wormhole link. Fig. 2 represents this case.

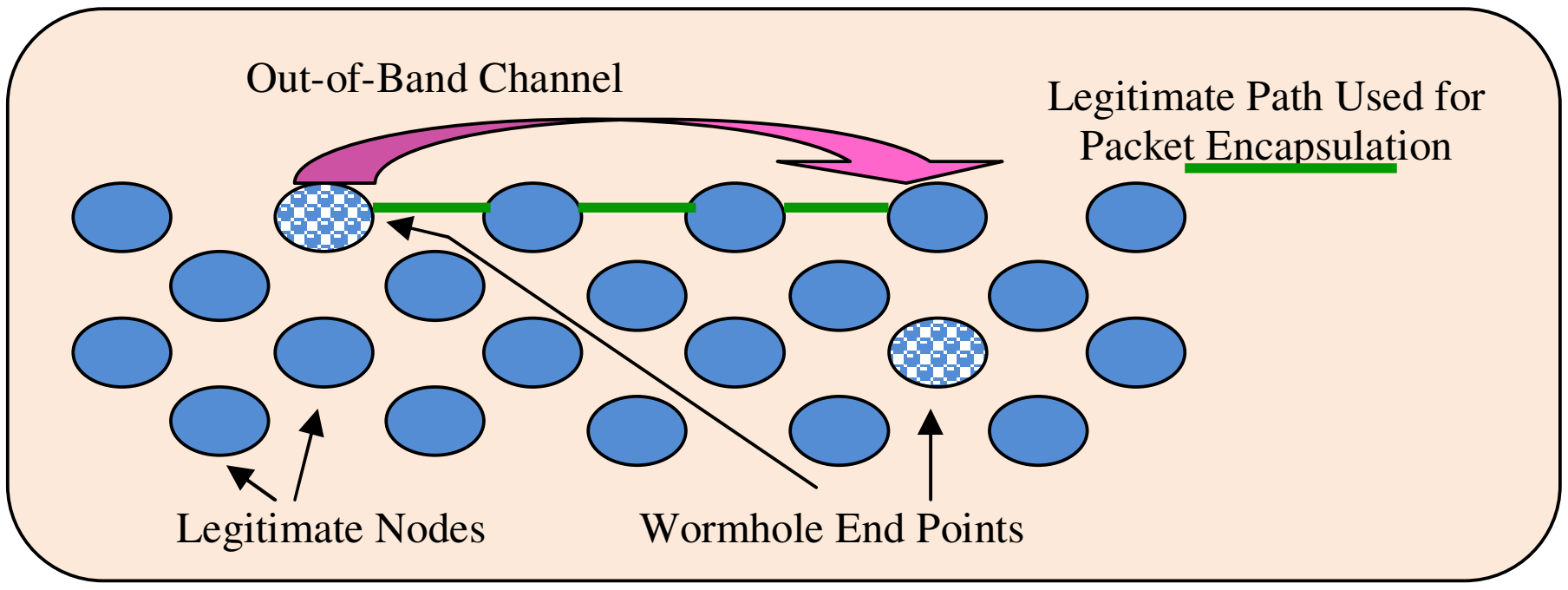

Figure 2. Wormhole using out-of-band channel and encapsulation

***Wormhole using Packet Encapsulation***

Each packet is routed via the legitimate path only, when received by the wormhole end, gets encapsulated to prevent nodes on way from incrementing hop counts.The packet is brought into original form by the second end point.

***Wormhole using High Power Transmission***

This kind of wormhole approach has only one malicious node with much high transmission capability that attracts the packets to follow path passing from it.

***Wormhole using Packet Relay***

Like the previous approach, only one malicious node is required that replays packets between two far nodes and this way fake neighbors are created.

***Wormhole using Protocol Deviation***

The malicious node creates wormhole by forwarding packets without backing off unlike a legitimate node and thus, increases the possibility of wormhole path getting selected. [5]

### 2.3.2. Models of Wormhole Attacks

Packet forwarding behavior of wormhole end points as well as their tendency to hide or show the identities, leads to the following three kinds of models. Here, S and D are the source and destination respectively. Nodes $M_1$ and $M_2$ are malicious nodes.

***Open Wormhole***

Source and destination nodes and wormhole ends $M_1$ and $M_2$ are visible. Identities of nodes A and B, on the traversed path are kept hidden.

***Half-Open Wormhole***

Malicious node $M_1$ near the source is visible, while second end $M_2$ is set hidden. This leads to path S-$M_1$-D for the packets sent by S for D.

***Close Wormhole***

Identities of all the intermediate nodes on path from S to D are kept hidden. This leads to a scenario where both source and destination feel themselves only one-hop away from each other. Thus fake neighbors are created. [6]

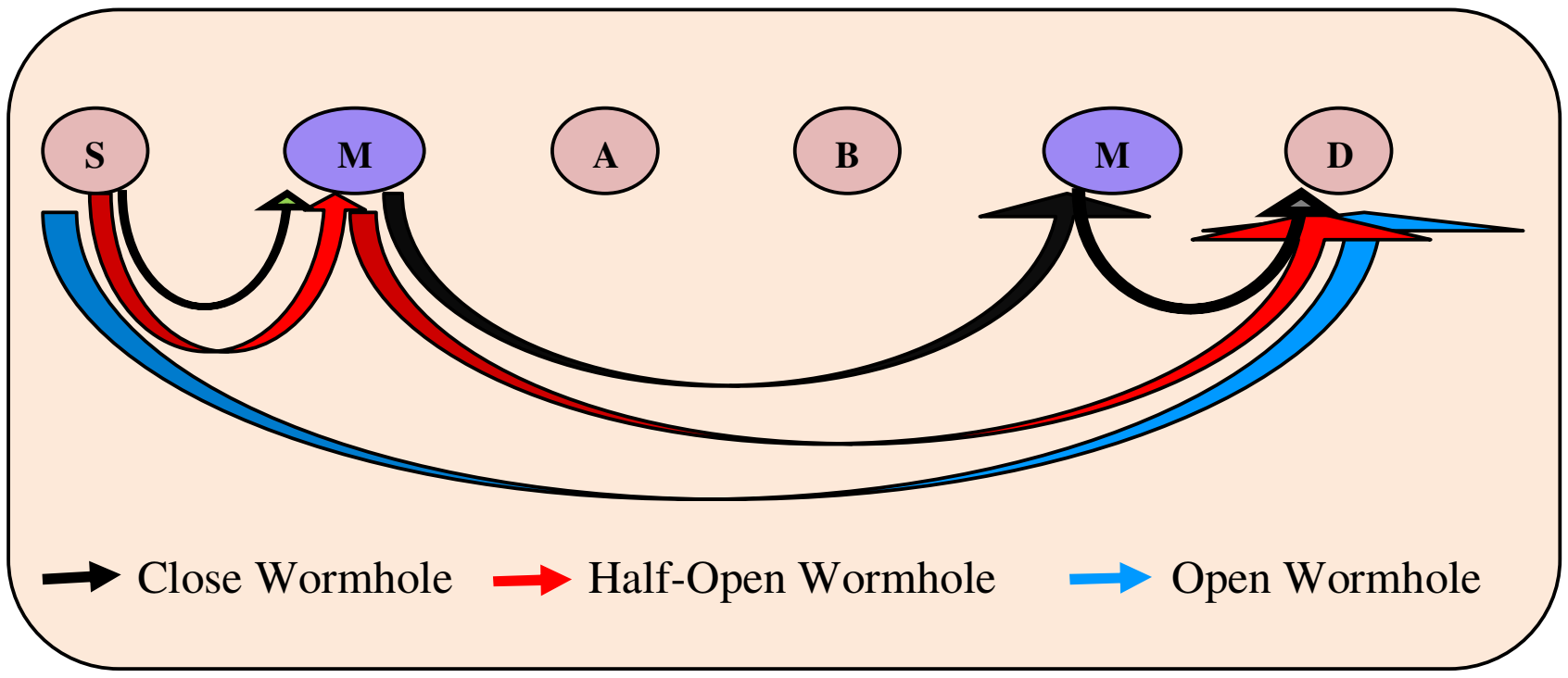

Figure 3. Open, Half-Open and Closed Wormhole

## 3. RELATED WORK

***Location and Time based approaches***

Hu and Perrig [7] presented an approach using Packet Leashes, where in geographic leash and temporal leash put upper bound on location of the receiver and maximum time a packet takes to travel respectively. TIK protocol is proposed for defense against temporal leash, but the knowledge of geographic location or tight time synchronization is required. Taheri, Naderi and Barekatain [39] used leashes approach with modified packet transmission methodology to decrease calculation overhead of TIK protocol.

In transmission time based mechanism (TTM), Tran, Hung and Lee brothers [8] proposed an approach where each node on path notes time of sending RREQ packet and receiving RREP packet. Here, also time consideration is the main factor. Singh and Vaisla [28] modified this approach by removing the sender and receiver from maintaining request and reply packet timing.

Hu and Evans [9] proposed a location based approach, where directional antenna is used to check the validity of neighbor. Considering the direction from which the response of HELLO message comes and using verifiers, the neighbors are authenticated. The approach can detect insider attack also by establishing authentication with pair wise secret keys, but hardware support is required here. Additionally, only types of wormholes with fake neighbors can be detected with this methodology.

Khalil, Bagchi and Shroff [5] proposed a lightweight countermeasure (LITEWORP) for wormhole attack detection using guard nodes. After detecting wormhole, LITEWORP leaves network in that open mode only, causing possibility of more disruption. To overcome this, they proposed another protocol MOBIWORP [10], which removes malicious nodes from the network using central authority either locally or globally.

Chen, Lou, Sun and Wang [16] presented a secure localization approach that can detect simplex and duplex wormhole attacks. They extended this algorithm [12] to make it effective for dissimilar transmission range of sensor nodes also, but still multiple wormholes cannot be detected by this.

Nait-Abdesselam, Bensaou and Taleb [13] proposed detection and avoidance method that focuses on the load carrying by various routes. When a route is loaded heavily, it may be because of packet congestion etc., so it may signal alarm even when wormhole is not present.

Khurana and Gupta [14][15] proposed an approach based on the travelled distance and maximum transmission range of sensor nodes. SEEEP [14] was limited to the nodes with same transmission range that has been extended as FEEPVR [15] to support dissimilar ranges also.

Jakob Eriksson, Srikanth V. Krishnamurthy, MichalisFaloutsos [16] presented TrueLink concept that has rendezvoused and authentication phases for wormhole detection. The former phase requires tight time synchronization, while the later works on shared secret keys for signing messages.

***Connectivity and Neighborhood based approaches***

Hayajneh, Krishnamurthy and Tipper [17] presented SECUreNeighborhooD (SECUND) protocol that can detect multi-ended wormhole. No need of specialized hardware, knowledge about the node locations and no requirement of clock synchronization are positive points of this method, but it can work only if the presence of wormhole increases fake neighbors by considerable amount.

Dimitriou and Giannetsos [18] derived an algorithm for wormhole detection based on connectivity information. The algorithm runs local path existence test when it detects new nodes.

Gupta, Kar and Dharmaraja [25] presented an approach where the presence of wormhole is found by the destination by counting hop difference between the neighbors of one hop away nodes. Special kind of Haund Packets is used for this purpose that introduces some processing delay also.

Vani and Rao [26] proposed an approach WARRDP (Wormhole-Avoidance Route Reply decision packet)for wormhole detection and removal using the combined approach of Hop count, Anomaly based and Neighbor list methods.

***Graphical and Topological Information based approaches***

Wang and Bhargava [19] presented a centralized approach MDS-VOW (Multi-Dimensional Scaling- Visualization Of Wormhole) with central controller. Here no hardware support is required, but it is less effective for sparse network.

A graph theoretic approach was presented by R. Poorvendram and Lazos [20] that provides necessary and sufficient conditions to detect and defend against wormhole attack. Specialized guard nodes, with high radio range, are the requirements of this methodology.

Choi, Kim, Lee and Jung [21] proposed a Wormhole Attack Prevention (WAP) algorithm based on DSR protocol. The algorithm works well for hidden attacks, but for the exposed attacks, it is difficult to detect by this approach.

Azer, Kassas and Soudani [22] proposed a detection and prevention approach based on Diffusion of Innovations that works fine except the end to end delivery time is increased considerably.

***Routing Algorithm Specific approaches***

Poornima, Bindu and Munwar [23] proposed a scheme based on geographic routing that, with Reverse Routing Scheme (RRS) and Authentication of Nodes Scheme (ANS), detects the presence of wormhole. It mainly works for BSR protocol and the value of witness threshold is too critical for the success of this approach.

Attir, Abdesselam, Brahim, Bensaou, Ben-Othman [24] proposed an approach using neighborhood detection and using W-Delay and appending additional information to the HELLO packet, detects wormhole. This method works, but it is limited to OLSR protocol only.

# 4. PROPOSED APPROACH

An approach for wormhole prevention is presented here where each node forwarding Route Reply RREP packet checks the validity of the two-hop neighbor node forwarding that packet. To accomplish the presented technique, a unique key derived based on all the two-hop neighbors is provided to each sensor node in the initial phase.

## 4.1.Key Generation Phase

This phase is carried out when a sensor network is established or an external node wishes to be a part of the network. The phase starts with execution of TinyPK [28] protocol to include each sensor node in the existing network.

The distribution of unique ID and Key value can be employed as explain below.

**Key Distribution**

LEAP [29] protocol proposes a technique to establish an individual key for each node shared with the base station. In that, a unique ID u is assumed to be assigned to each sensor node. By applying some pseudo random function f to the unique ID with the master key $K_m$, a unique main key$K_{mu}$ is generated as

$$K_{mu} = f(K_m, u)$$

Here, the master key is common and available only to the controller. The generated key is pre-loaded into each node prior to its deployment.

To make this protocol applicable for the proposed approach, separate master keys for individual node is required to be used and the function f is designed to apply to all two-hop neighbor IDs in place of unique ID u. Additionally, information regarding the master key value $K_m$, generated key $K_{mu}$ and applied function f are also provided to each sensor node for further use.

## 4.2.Wormhole Prevention Phase

In commonly used Dynamic Routing Algorithms, Route Request (RREQ) packet is broadcasted by the source node. All nodes receiving this packet broadcast it further until it reaches to the destination. As shown in the Fig. 4, nodes A and O are source and destination nodes respectively. Node A is broadcasting RREQ packet. On receiving this packet, node O forwards Route Reply (RREP) packet for the path from which it obtains the first RREQ packet.

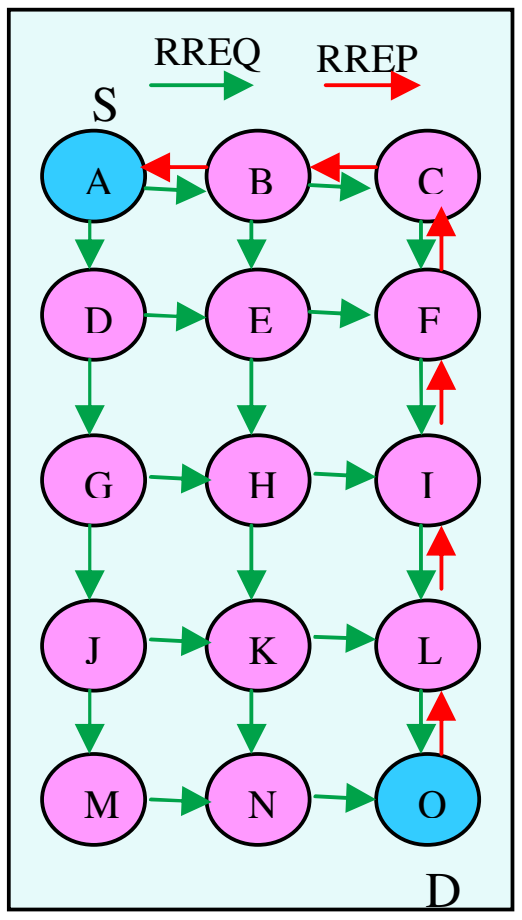

Figure 4. Transmission of RREQ and RREP Messages

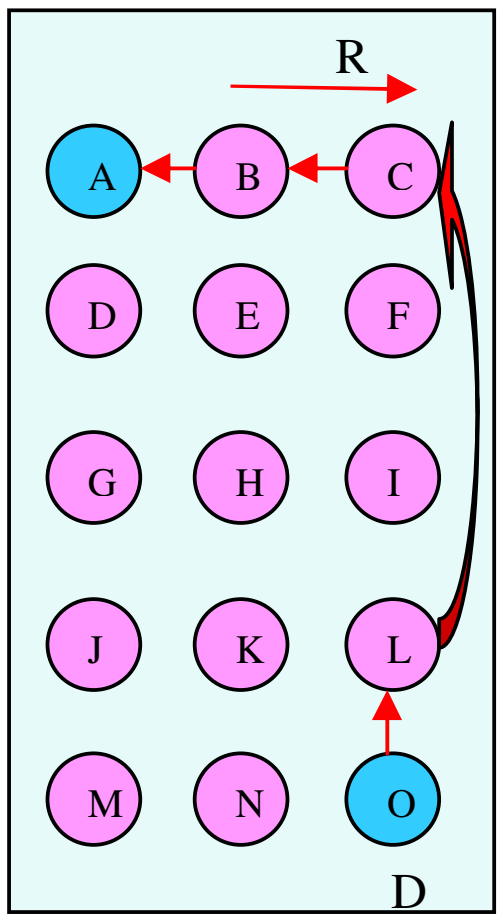

Figure 5. Transmission of RREP Message via Wormhole Link

Let use consider a case where a wormhole link is present between node C and node L. When RREQ is received by node C, it will be diverted to node L directly via the established out-of-band wormhole link. In this case, RREP packet follows the path shown in fig 5.

For wormhole prevention, each node is supposed to store the detail of each and every RREP packet it forwards. On receiving RREP, its validity is tested through a check phase which is started with broadcasting of *Probe* message and its corresponding *Probe_Ack_Tag* value. For various possible cases the sequence is explained further.

### 4.2.1. Case with No Wormhole

Let us for the above scenario, shown in fig. 4, no wormhole is present. The steps taken by the node on reception of RREP packet are explained here. If a node receives RREP from the destination,then forwards the packet else starts the following steps. The steps are explained with reference to node I.

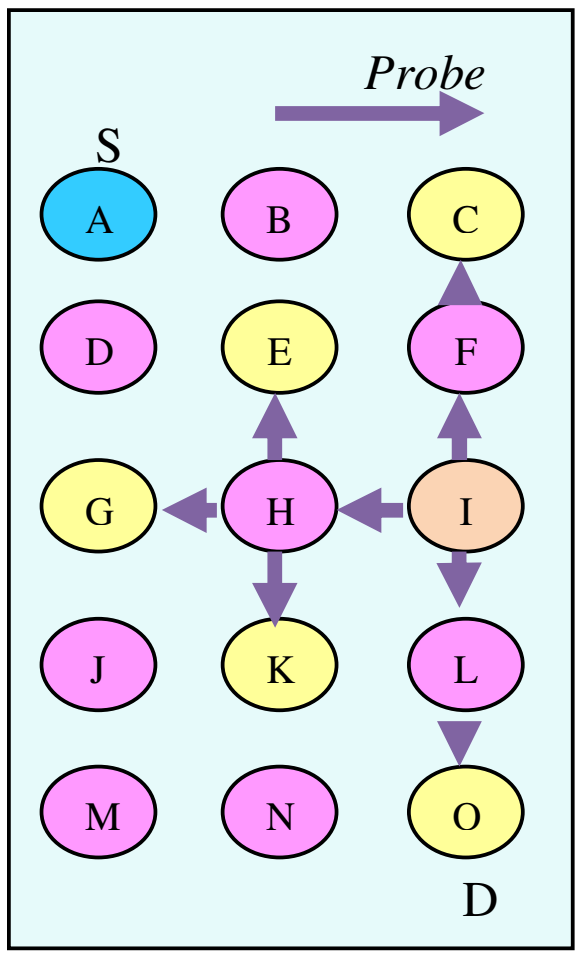

Figure 6. Transmission of *Probe* Message by Node I

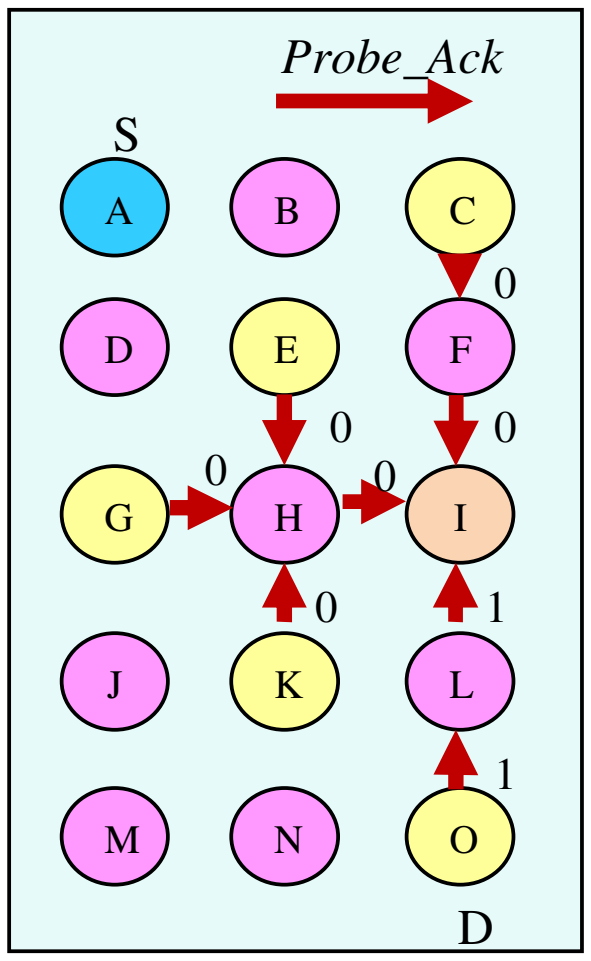

Figure 7. Transmission of *Probe_Ack* Message to Node I

1. Node I sends *Probe* message to all its two-hop neighbors i.e. to nodes C, E, G, K and O, where source and destination IDs A and O respectivelyare also specified as in Fig. 6
2. Each of the node checks whether it has sent or forwarded RREP for given source-destination combination and responds with *Probe_Ack* message accordingly
3. If has forwarded the corresponding RREP packet, then it attaches *Probe_Ack_tag* 1 with the *Probe_Ack* message else attaches 0 and sends back to node I
4. Let us assume that node I receives *Probe_Ack* messages as :
   C with 0, E with 0, G with 0, K with 0 and O with 1
5. Exactly one tag is 1 so the case is considered to be a valid case
6. Derived Local Key LK using node IDs for C,E,G,K and O and if it matches with its own $K_{mu}$, then forwards RREP message, else alarms "Illegal Case"
7. In general, for node N,

   If, $\sum_{i=1}^{N_{2hn}} Probe_Ack_Tag_i = 1$

   Where, $N_{2hn}$ = Two-hop Neighbors of Node N

   Then, calculate

$$LK(N) = f(ID_i, 1 \leq i \leq N_{2hn})$$

   If $K_{mu}(N) = LK(N)$, the RREP packet had arrived from a valid node and will be alarms "Illegal Case"

   forwarded,else

   Same Procedure is repeated by nodes F, C, B and A

### 4.2.2. Case with Wormhole

Let us assume a wormhole link is present between nodes C and L, so RREP is forwarded from node L to node C directly as shown in Fig. 5. For this, RREP gets forwarded till node B. Then node B sends *Probe* message. The various possible cases how the legitimate and fake neighbors may respond to the *Probe* message are discussed further.

#### 4.2.2.1. Possibility 1 : Hidden Passive Wormhole Attack

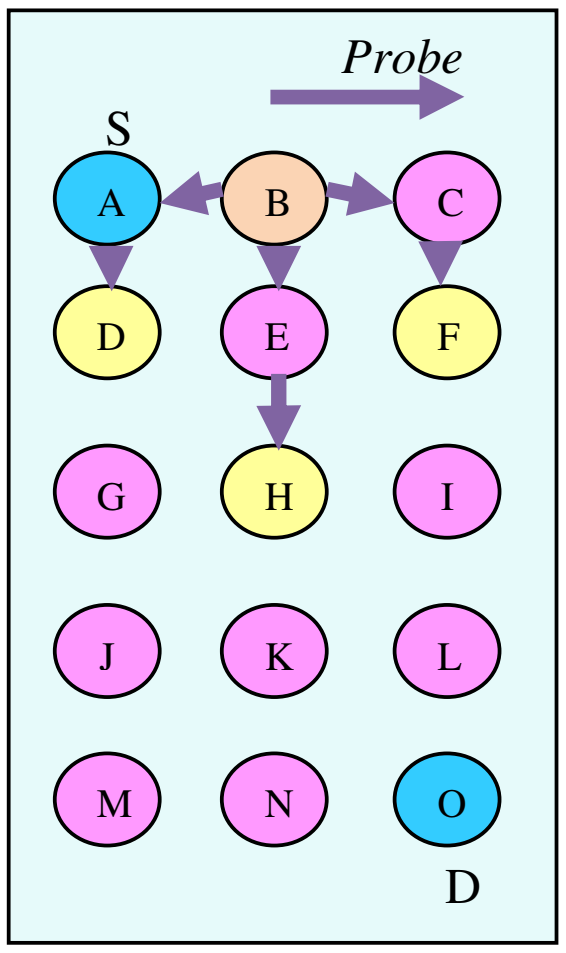

Figure 8 Transmission of *Probe* Message by Node B

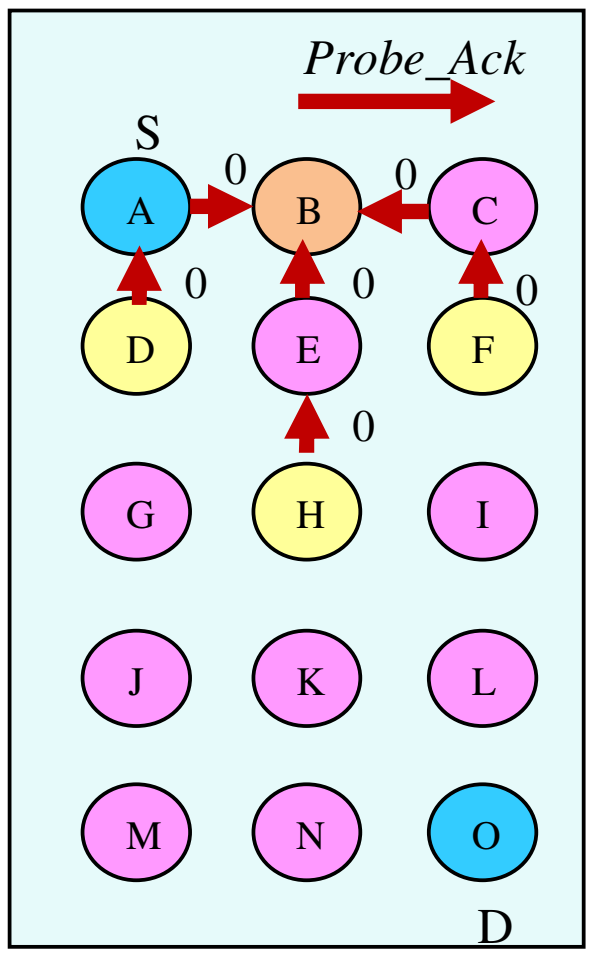

Figure 9. Transmission of *Probe_Ack* Message to Node B

As Shown in Fig. 9, assume that node B receives *Probe_Ack* messages as :D with 0, F with 0 and H with 0. Ideally, the situation is impossible where none of the two-hop neighbor node has forwarded the RREP packet. Identifying such results, node B alarms the "Illegal Case".

Let us for node N,

If, $$\sum_{i=1}^{N_{2hn}} Probe_Ack_Tag_i = 0$$

then, the case is invalid case and RREP is stopped getting forwarded.

#### 4.2.2.2. Possibility 2 : Exposed Passive Wormhole Attack

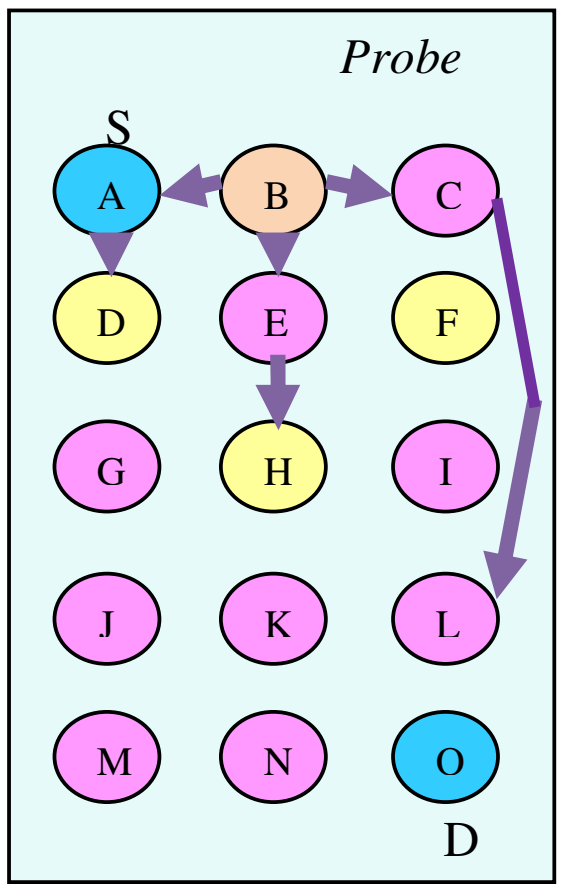

Figure 10. Transmission of *Probe* Message by Node B

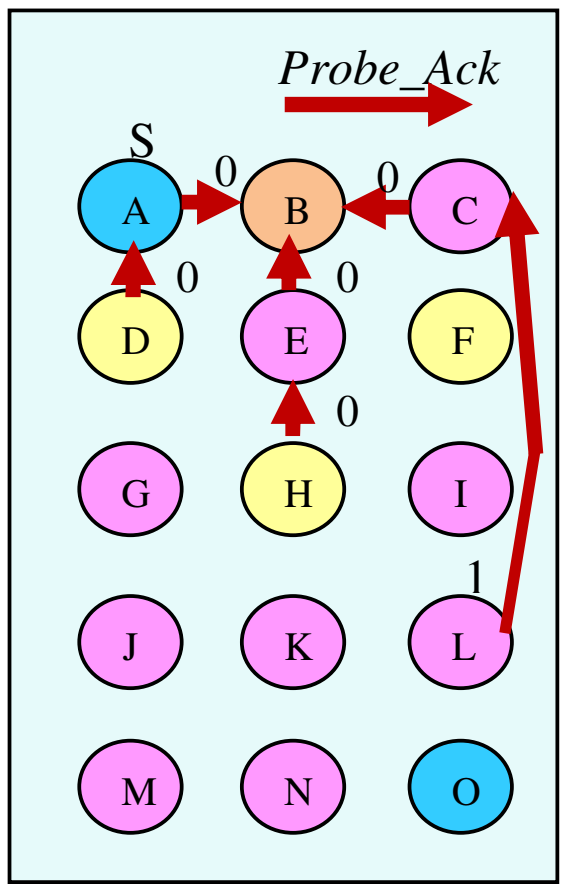

Figure 11. Transmission of *Probe_Ack* Message to Node B

Let us consider a case where wormhole end L keeps its identity exposed and node B receives *Probe_Ack* messages as :D with 0, F with 0, H with 0 and L with 1. Here, when LK is derived considering the IDs of nodes D, F, H and L, its value does not match with the $K_{mu}$ value of node B as node L is not the legitimate two-hop neighbor and has not considered while deriving $K_{mu}$.

In general, for node N,

If, $$\sum_{i=1}^{N_{2hn}} Probe_Ack_Tag_i = 1$$

then, calculate

$$LK(N) = f(ID_i, 1 \leq i \leq N_{2hn})$$

Here, $K_{mu}(N)$ <> LK(N), as the RREP packet has arrived from a malicious node pretending as a two-hop neighbor node. Accordingly, this case is also alarmed as an "Illegal Case".

#### 4.2.2.3. Possibility 3 : Hidden Active Wormhole Attack

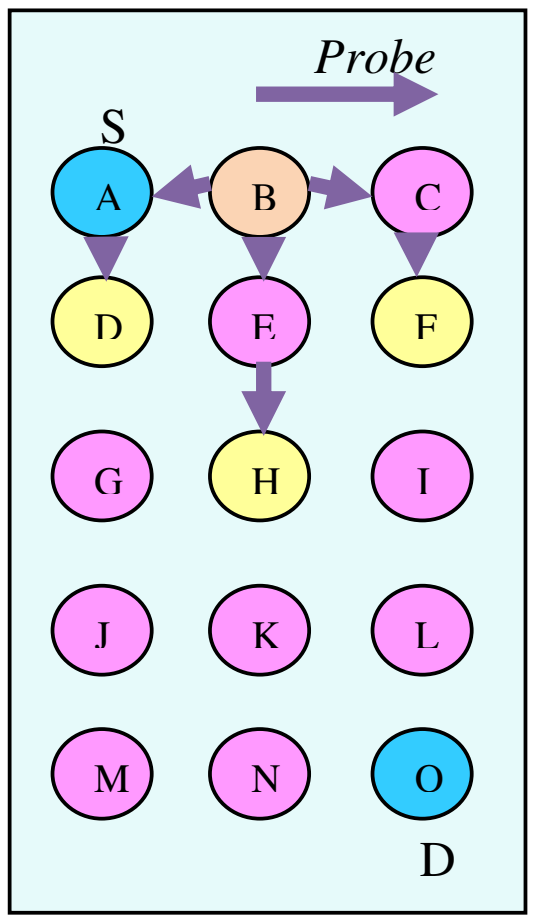

Figure 12. Transmission of *Probe* Message by Node B

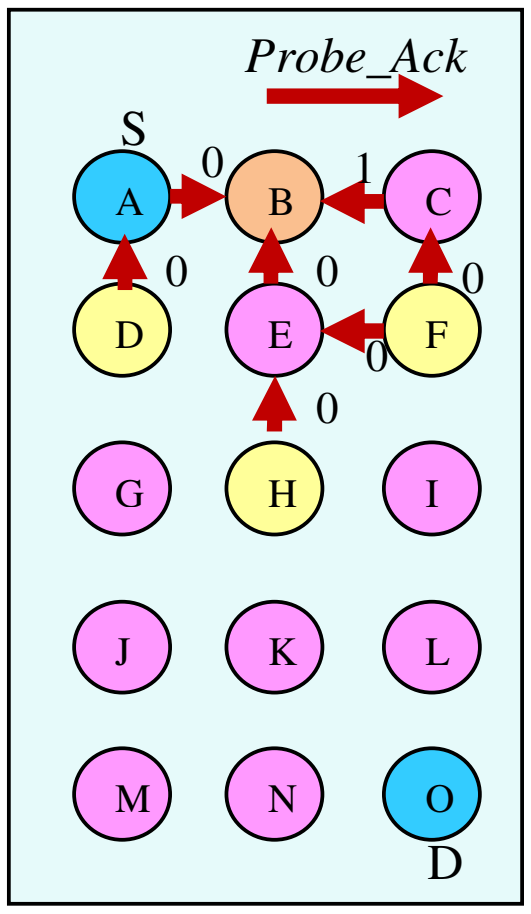

Figure 13. Transmission of *Probe_Ack* Message to Node B

A case with active attack can be considered where node F sends *Probe_Ack_Tag* with 0 value, but it is changed to 1 by the malicious node C. In such case the received*Probe_Ack* messages are:  D with 0, F with 1 and H with 0.

This kind of result finds matching between the derived LK value and Kmu value and so validates the two-hop neighbor sender and forwards the RREP packet. In reality, the case is an illegal case as maliciously the tag value sent by one of the legitimate node is altered on way. However, if the sent *Probe_Ack* message by node F reaches to node B for more than once via different routes, then node B received dissimilar tag values for the node F. Such situation raises a question on the validity of received tag values and "Illegal Case" is alarmed.

Let, $Probe_Ack_Tag_{ij}$ = Tag Message Received by node N from node j, sent by node i
Then, for the defined case, if we take N=B,
$Probe_Ack_Tag_{FC}$= 1 and $Probe_Ack_Tag_{FE}$= 0, where ideally, both must be equal.

i.e. if for a node N, if $\sum_{i=1}^{N_{1hn}} Probe_Ack_Tag_{ij} \leq 1$

then, for $N_{1hn}$ = One-Hop Neighbor of node N, the tag message is likely to be altered on way. Ifj=1, it meansthat there is only one common one-hop neighbor between node N and its two-hop neighbor I ,then we may not be able to find that the *Probe_Ack_Tag*valuehas been changed on way.

## 5. SIMULATION SETUP

Simulations are performed in ns-2 network simulator [30] for 30 nodes, where parameters are given in Table 1.  Zero background traffic and nodes with zero mobility are assumed here.

Table 1: Simulation Parameters

| Parameters | Values |
|---|---|
| Examined Protocol | AODV |
| Simulation time | 0.3 units |
| Simulation area | 600 × 600 |

| Parameters | Values |
|---|---|
| Transmission range | 250 m |
| Traffic type | CBR(UDP) |
| Number of wormholes | 3 |

# 6. RESULTS AND ANALYSIS

The proposed solution allows sensor nodes to forward RREP packet provided some conditions are met. This gets accomplished by spending additional time and energy. An upper limit can also be placed on waiting time for *Probe_Ack* messages that is also decided using simulation.

## 6.1. RREP Transmission Time and Energy Consumption

Before forwarding the RREP packet, each node broadcasts Probe message and waits for *Probe_Ack* message from the two-hop neighbor nodes. Based on the decision taken from the received tag values, decision for forwarding RREP is taken. These additional steps add total transmission time in receiving the RREP packet as well as the energy consumed by the nodes. The graphs shown in fig. 14 and fig. 15compare the transmission time taken and energy consumed with and without the implementation of the proposed prevention approach respectively.

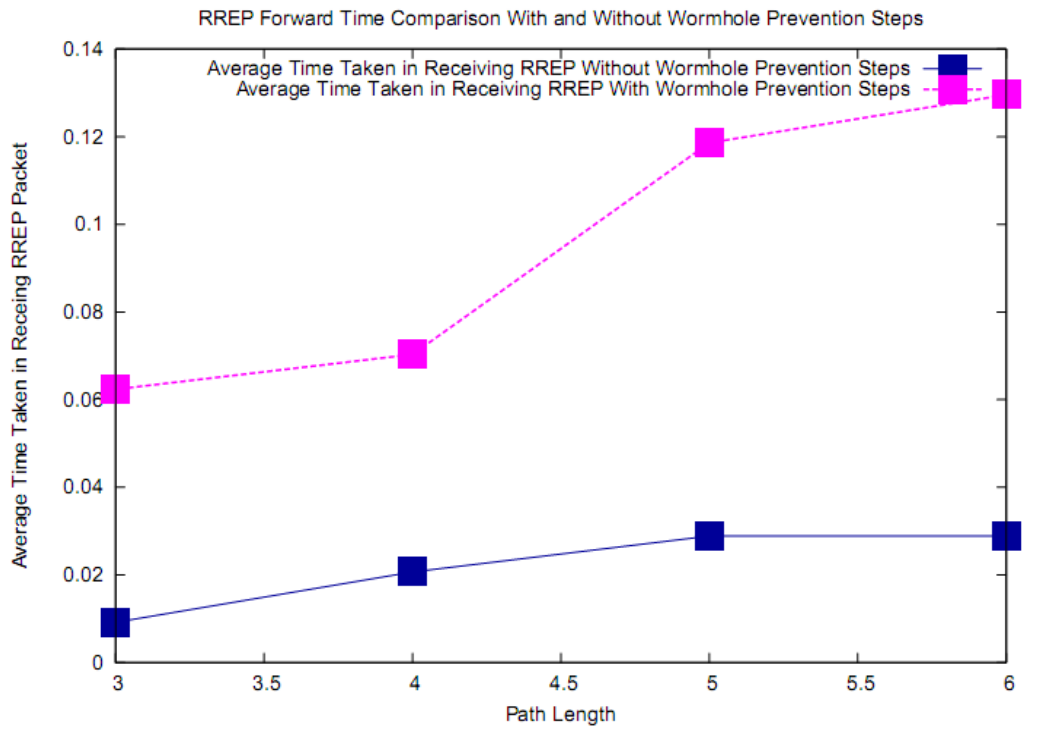

Figure 14. Transmission Time Taken in RREP Packet

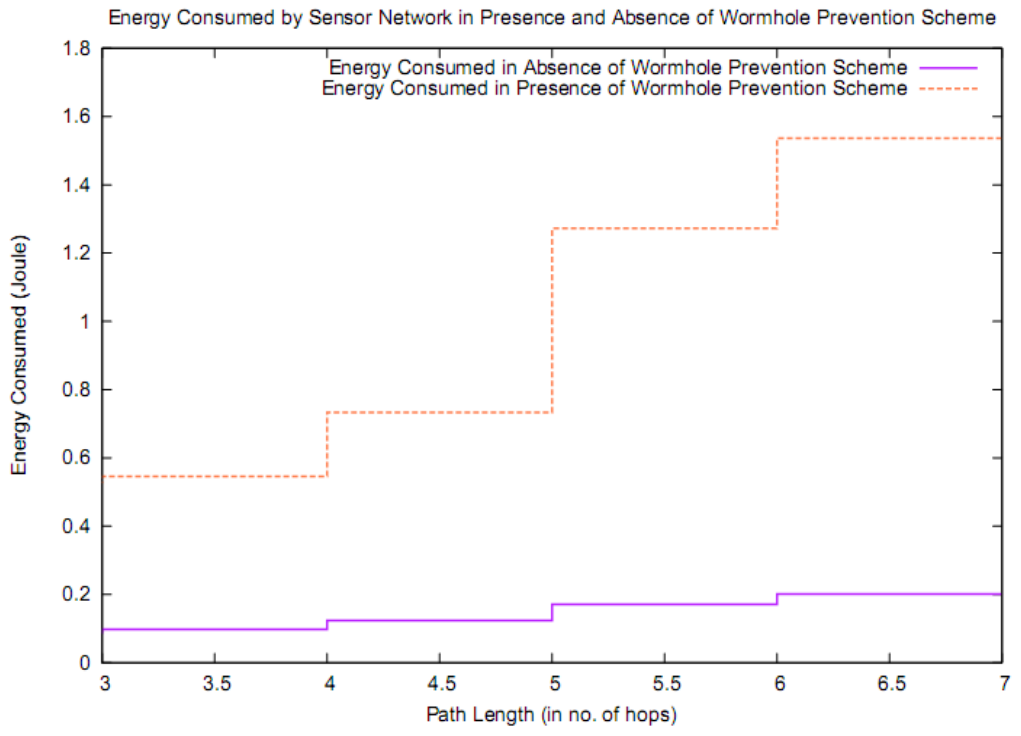

**Figure 15.** Node Energy Consumption during RREP Packet Transmission

### 6.2. Time Taken in Collecting *Probe_Ack*Messages

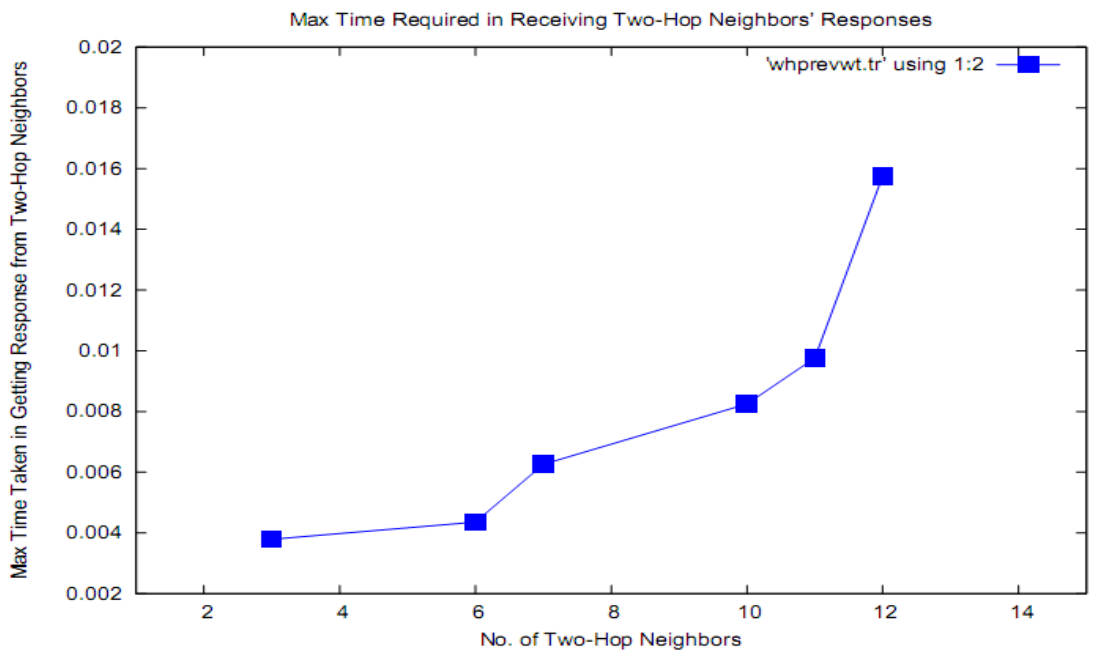

Figure 16. Time Consumption in Collecting *Probe_Ack*

The graph in fig. 16plots time taken in receiving the *Probe_Ack* message. It shows that after a node broadcasts *Probe* message, within maximum1 time unit, it receives all of the *Probe_Ack* messages, so that can be taken as the maximum time a node can wait for checking the validity of RREP packet.

## 7. Conclusions and Future Scope

The approach proposed here makes RREP packet forwarding conditional. By checking the validity of the two-hop neighbor node that has forwarded the packet, a node lets it to move further towards the source. Wormhole end is detected when the identity of the two-hop neighbor is found illegal. Authenticity checking of such two-hop neighbors is carried out using a preloaded secret key. By comparing the memory requirement for various numbers of neighbors, it can be concluded that by spending more on setup cost, higher scalability can be achieved. The proposed scheme focuses on the type of wormhole with out-of-band channel. It can be extended to detect other types of wormhole attacks also.

## Authors

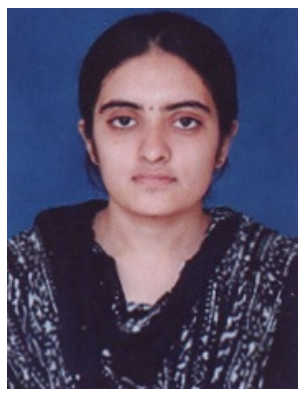

**Dhara Buch** is serving as an Assistant Professor in Department of Computer Engineeringat Government Engineering College, Rajkot (India). She is also pursuing her M. Tech (R) from S. V. National Institute of Technology, Surat. Her area of interest includes Information Security in Wireless Sensor Networks.

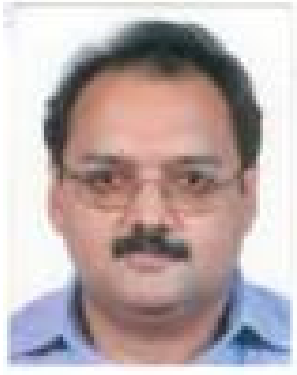

**Dr. DeveshJinwala**is serving as an Associate Professor in Computer Engineering with Sardar Vallabhbhai National Institute of Technology, Surat (India). His research work is on Configurable Link layer Security Protocols for Wireless Sensor Networks. His major areas of interest are Information Security Issues in Resource Constrained Environment, Algorithms & Computational Complexity and Software Engineering.